RESEARCH ARTICLE                                                                OPEN ⓐ ACCESS

# GPS scintillations associated with cusp dynamics and polar cap patches

Yaqi Jin[1,*], Jøran I. Moen[1,2], Kjellmar Oksavik[2,3,4], Andres Spicher[1], Lasse B.N. Clausen[1] and Wojciech J. Miloch[1]

[1] Department of Physics, University of Oslo, P.O. Box 1048 Blindern, 0316 Oslo, Norway
[2] Arctic Geophysics, University Centre in Svalbard, N-9171 Longyearbyen, Norway
[3] Birkeland Centre for Space Science, Department of Physics and Technology, University of Bergen, P.B. 7803, 5020 Bergen, Norway
[4] Now visiting Center for Space Science and Engineering Research (Space@VT), Virginia Tech, 1901 Innovation Drive, Blacksburg, VA 24060, USA

**Abstract** – This paper investigates the relative scintillation level associated with cusp dynamics (including precipitation, flow shears, etc.) with and without the formation of polar cap patches around the cusp inflow region by the EISCAT Svalbard radar (ESR) and two GPS scintillation receivers. A series of polar cap patches were observed by the ESR between 8:40 and 10:20 UT on December 3, 2011. The polar cap patches combined with the auroral dynamics were associated with a significantly higher GPS phase scintillation level (up to 0.6 rad) than those observed for the other two alternatives, i.e., cusp dynamics without polar cap patches, and polar cap patches without cusp aurora. The cusp auroral dynamics without plasma patches were indeed related to GPS phase scintillations at a moderate level (up to 0.3 rad). The polar cap patches away from the active cusp were associated with sporadic and moderate GPS phase scintillations (up to 0.2 rad). The main conclusion is that the worst global navigation satellite system space weather events on the dayside occur when polar cap patches enter the polar cap and are subject to particle precipitation and flow shears, which is analogous to the nightside when polar cap patches exit the polar cap and enter the auroral oval.

## 1 Introduction

The ionospheric electron density irregularities can modify trans-ionospheric radio waves, which results in rapid fluctuations of the received amplitude and phase on the ground, known as ionospheric scintillations (see e.g., Yeh & Liu, 1982; Kintner et al., 2007, and references therein). The ionospheric scintillation can strongly affect the satellite based navigation system, such as global navigation satellite system (GNSS), and high frequency (HF) communication systems (Carlson, 2012). Due to the practical reasons, the GPS scintillation studies have received more and more attention at high latitudes recently (e. g., Mitchell et al., 2005; Spogli et al., 2009; Prikryl et al., 2010; Alfonsi et al., 2011; Jin et al., 2014; van der Meeren et al., 2014; Wang et al., 2016).

It was shown by Jin et al. (2014) and van der Meeren et al. (2015) that the strongest GPS phase scintillations at high latitudes occur when the polar cap patches, islands of enhanced F region ionosphere, exit the polar cap into the nightside auroral region. This finding was further confirmed by a comprehensive statistical study (Jin et al., 2016). In order to distinguish the large-scale plasma structures between different sources (solar EUV versus auroral impact), Jin et al. (2016) defined two types of auroral blobs at the nightside sector: blob type 1 (BT-1) which is due to a polar cap patch that has entered the auroral oval, and blob type-2 (BT-2) which is due to the auroral dynamics alone. Their results show that the BT-1 blobs are associated with the highest scintillation level. However, the question whether this knowledge can be extrapolated to the dayside sector has remained open. On the dayside, the cusp ionosphere appears to be an active region in terms of GPS scintillation (Moen et al., 2013), where intense scintillation and loss of signal lock are known to occur (Oksavik et al., 2015). By analyzing the interplanetary magnetic field (IMF) $B_y$ dependence of the GPS phase scintillation around magnetic noon, Jin et al. (2015) suggested that GPS phase scintillations are sensitive to a combination of the cusp aurora and the intake of solar EUV-ionized plasma. This paper is devoted to test this suggestion.

In this paper, we present GPS scintillations in the European Arctic sector near the dayside cusp inflow region around magnetic noon. On December 3, 2011, the EISCAT Svalbard radar (ESR) observed modest plasma density in the prenoon convection cell and well structured high density plasma (polar cap patches) in the postnoon convection cell. The observed high density plasma were transported from the subauroral region. The different plasma conditions from prenoon and postnoon gives us an opportunity to compare the scintillation levels associated with the two different ionospheric conditions. The flow pattern is monitored by SuperDARN and the ionospheric plasma is characterized in detail by the ESR. The result shows that the combination of cusp dynamics and the formation of polar cap patches produce a significantly higher GPS phase scintillation level than cusp dynamics without the formation of polar cap patches.

*Corresponding author: yaqi.jin@fys.uio.no

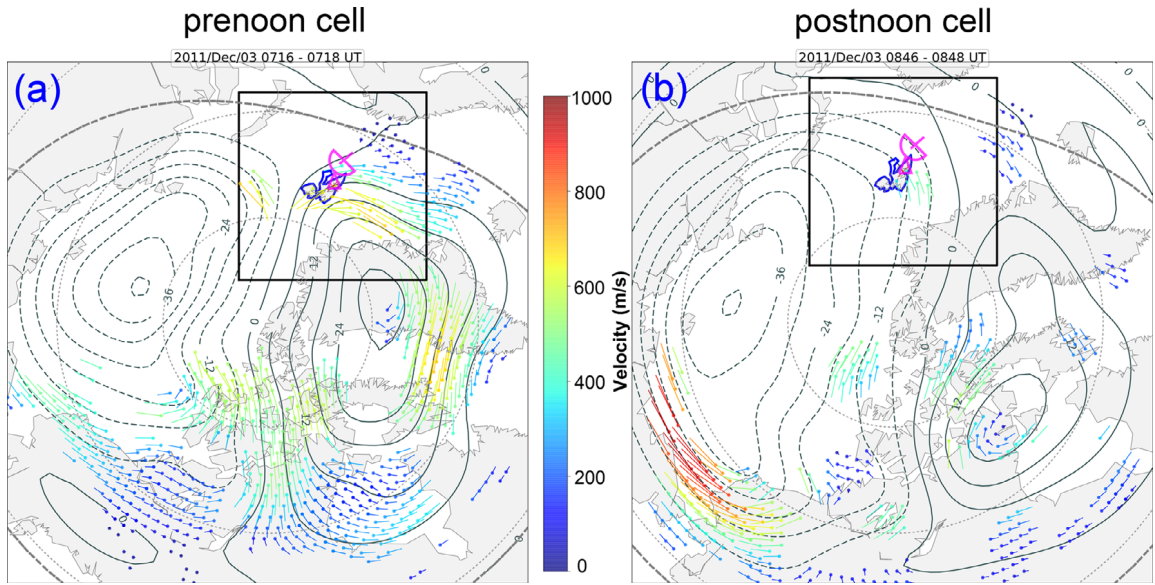

**Fig. 1.** Fitted drift velocities and the electric potential from SuperDARN showing the twin-cell convection pattern. The location of the ESR is highlighted by a radar symbol in a rectangle box to show the observations in the prenoon (a) and postnoon (b) convection cells, respectively.

## 2 Instrumentation

In this study, we use the GPS scintillation receivers at Ny-Ålesund (NYA) and Longyearbyen (LYB). The NYA receiver was operated by the University of Oslo (UiO), while the LYB receiver was operated by the Istituto Nazionale di Geosica e Vulcanologia (INGV, Italy) (Romano et al., 2008, 2013). Both are the standard NovAtel GSV4004 GPS ionospheric scintillation and TEC monitors (GISTM) which can record GPS total electron content (TEC) based on dual frequency measurements. The amplitude ($S_4$) and phase scintillation indices ($\sigma_\phi$) are also recorded based on 50 Hz measurements of the power and carrier phase at the GPS L1 frequency (1.57542 GHz) (Van Dierendonck et al., 1993). In order to minimize the multi-path effect, a cutoff elevation angle of 20° is used for the GPS TEC and scintillation data from Ny-Ålesund and GPS TEC and phase scintillation data from Longyearbyen, while a cutoff elevation angle of 30° for the amplitude scintillation data from Longyearbyen.

The ESR at Longyearbyen (magnetic noon is at 8:50 universal time (UT)) includes a 42 m antenna which is fixed to magnetic field-aligned (azimuth = 184°, elevation = 82°) and one 32 m antenna that is fully steerable (Wannberg et al., 1997). Only data from the 42 m antenna are used in this paper.

The SuperDARN data are retrieved from Virginia Tech using the DaViTpy software package. SuperDARN convection patterns were created using the technique of Ruohoniemi & Baker (1998) which uses data from all available superDARN radars. It generates the flow patterns every 2 min based on a fit of the ionospheric electrostatic potential in spherical harmonics (Greenwald et al., 1995; Chisham et al., 2007).

We also use observations of electrons and protons from the total energy detector (TED) and the medium energy proton and electron detector (MEPED) onboard NOAA-16 and NOAA-17 (Evans & Greer, 2000). The NOAA satellites are operated by the National Oceanic and Atmospheric Administration in low-altitude circular polar orbits. NOAA-16 and NOAA-17 are operated at altitudes of 861 and 822 km, respectively. The electrons and protons in 16 energy channels between 0.05 and 20 keV are measured by the TED in two looking directions (toward local zenith and 30° from zenith). MEPED measures energetic electrons and protons at angles of 10° and 80° to the local zenith, which at high latitudes correspond to precipitating and trapped particles, respectively. MEPED measures protons at 3 energy channels of 30–80, 80–250, and 250–800 keV, while it measures electrons at 3 energy channels of >30, >100, and >300 keV. We only use the electron data in this study.

The all-sky imager is often used in studies of polar cap patches and GNSS scintillations (Moen et al., 2006; Jin et al., 2014, 2015; Oksavik et al., 2015; Hosokawa et al., 2016). The University of Oslo all-sky imagers at Ny-Ålesund and Longyearbyen were operated during the time period of interest. However, due to the unfavorable cloud coverage above both stations, the data could not be used in this study.

## 3 Observations

To illustrate the high latitude flow pattern in the global context, we present in Figure 1a and b the SuperDARN

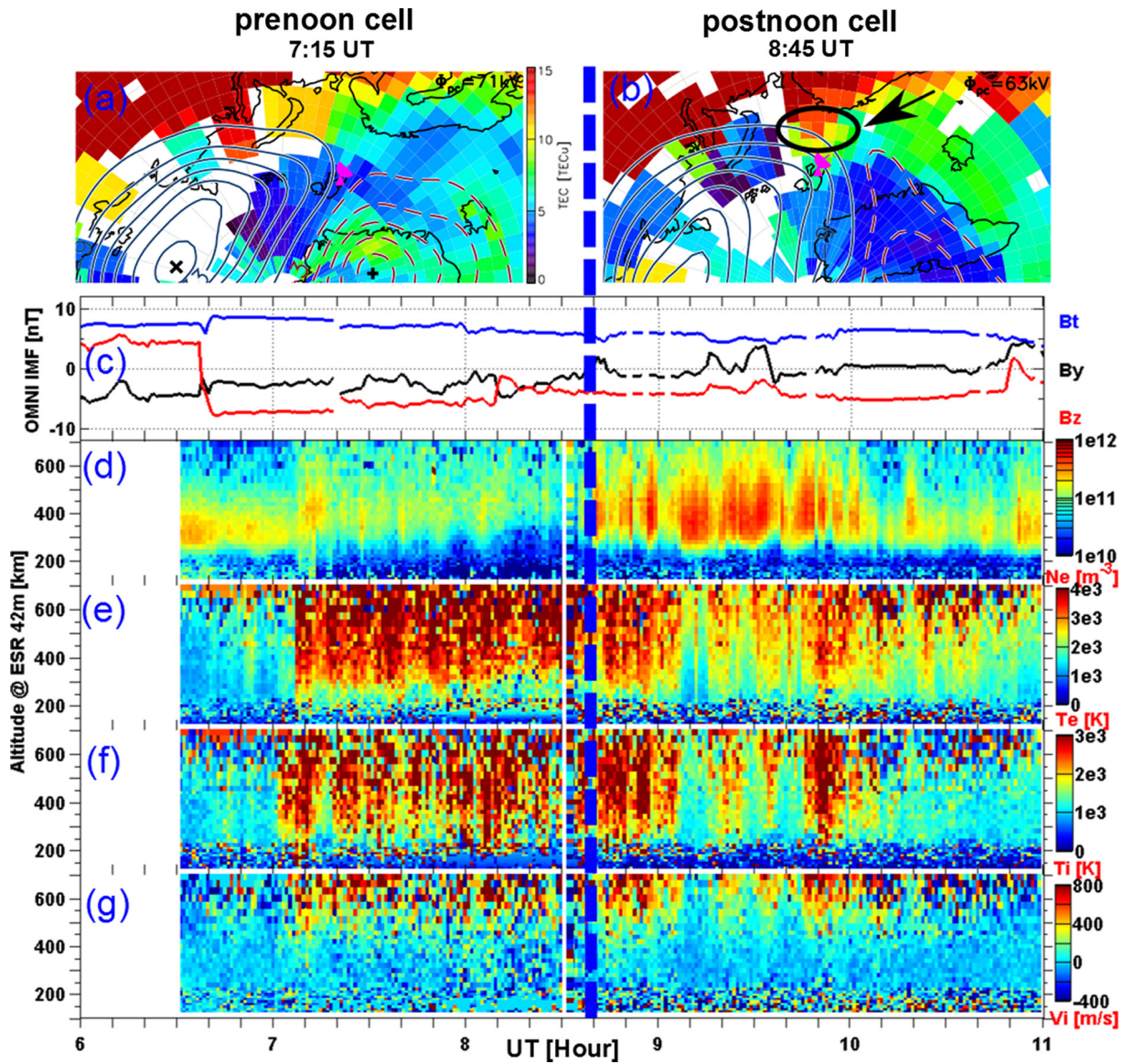

**Fig. 2.** Event overview on December 3, 2011. (a, b) The GPS TEC maps at 7:15 UT and 8:45 UT, respectively. The ESR location is indicated by a pink radar symbol above Svalbard. (b) The high density plasma moved poleward toward the Svalbard archipelago and it is annotated by a black circle and a black arrow. (c) The total magnetic field (Bt), By and Bz component of IMF from the OMNI dataset. (d–g) The electron density (Ne), electron temperature (Te), ion temperature (Ti), and line of sight ion velocity (Vi) as measured by the ESR 42 m antenna at Longyearbyen. The vertical dashed blue line delimits the prenoon and postnoon sectors.

convection maps at 7:16 UT and 8:46 UT on December 3, 2011, respectively. The Svalbard archipelago was in the prenoon convection cell before 8:40 UT. The ESR 32 m antenna was operated in an azimuthal scan mode from 6:30 UT to 8:30 UT, and several reversed flow channels (RFEs) (Rinne et al., 2007) were observed. The plasma that entered into the polar cap in the prenoon convection cell was characterized by modest electron density which was transported from the subauroral region from the prenoon sector (see Fig. 2 for more details). The lack of high density indicates that no polar cap patches were formed in the prenoon sector. However, other cusp phenomena such as particle precipitation and flow shears (RFEs) were observed (Spicher et al., 2016).

As the Earth rotated, the Svalbard archipelago moved to the postnoon convection cell after 8:40 UT which allows the access to higher density plasma from the postnoon sector. Indeed, the ESR 42 m antenna observed a series of high density cold plasma after 8:40 UT. These were the newly formed polar cap patches. The obvious different conditions, with and without polar cap patches, in the prenoon and postnoon convection cells allow us to characterize the GPS scintillations for the two different cases.

In order to illustrate the different ionospheric conditions when Svalbard was in the prenoon and postnoon convection cells, we show the combined GPS TEC map the convection pattern in Figure 2a and b. The TEC data were obtained through the Madrigal database. Figure 2a shows that the plasma density (in terms of TEC) around Svalbard was low when it was in the prenoon convection cell. There was no intake of the high density plasma into the polar cap. When

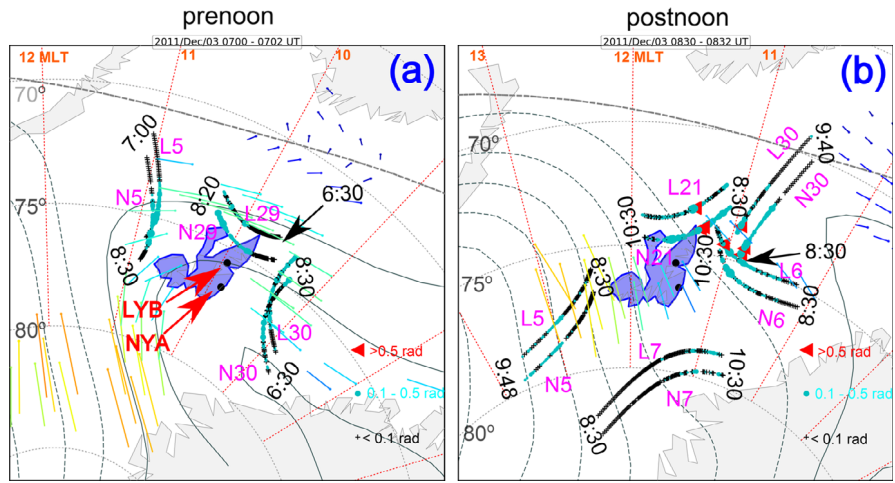

**Fig. 3.** The ionospheric convection pattern and locations of selected GPS satellites which were tracked at Ny-Ålesund (NYA) and Longyearbyen (LYB) stations in the prenoon (a) and postnoon (b) convection cells. The data are projected onto a geomagnetic coordinate grid. The GPS satellites are projected onto the map by assuming an ionospheric pierce point (IPP) altitude of 350 km. The start and end time of each track are indicated by "HH:MM" in UT near the start and end of each track. The GPS IPPs are shown using different symbols and colors: black crosses indicate $\sigma_\phi < 0.1$ rad, cyan dots indicate $\sigma_\phi$ between 0.1 and 0.5 rad whose sizes are according to the GPS phase scintillation level, and red triangles indicate $\sigma_\phi > 0.5$ rad. The tracks of different GPS satellites are annotated by the receiving station and their PRN (pseudorandom noise) code. For example N5 indicates PRN5 tracked at NYA station, while L5 indicates PRN5 tracked at LYB station. The locations of NYA and LYB are shown by black dots which are annotated in panel a.

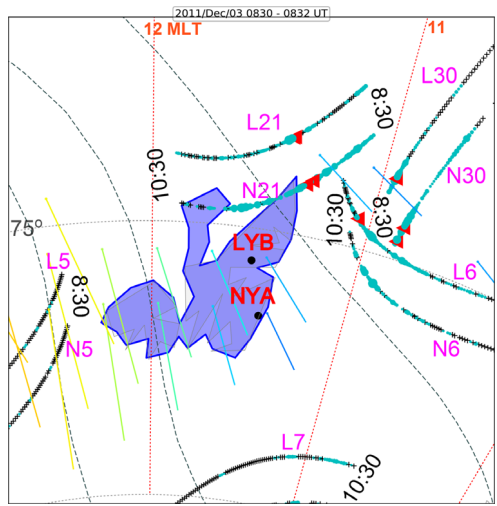

**Fig. 4.** The same format as Figure 3b but enlarged to better display the relative locations of the ESR and GPS IPPs.

Svalbard moved into the postnoon convection cell, the plasma density around it increased as the solar EUV-ionized plasma convected poleward from the postnoon subauroral sector to form polar cap patches around the cusp inflow region. In Figure 2b the black circle and black arrow mark a high density plasma structure which moved poleward toward Svalbard.

For the upstream IMF condition, we show in Figure 2c the OMNI data which were already time shifted from the measurement point to the Earth's bow shock nose (King & Papitashvili, 2005). The total magnetic field ($Bt$) decreased gradually from 7 nT at 6:00 UT to 5 nT at 11:00 UT. The IMF $Bz$ turned southward at 6:37 UT and remained negative in the following 4 h. The IMF $By$ was $-5$ nT at 6:00 UT and steadily increased until 8:40 UT, after which $By$ remained nearly 0 nT. The magnitude of $Bz$ was larger than $By$ during most of the period ($|Bz| > |By|$).

Figure 2d–g shows the ionospheric plasma parameters (electron density Ne, electron temperature Te, ion temperature Ti, and line of sight ion velocity Vi) from the ESR 42 m antenna. Due to the northward IMF condition before 6:37 UT, the open/closed magnetic field line boundary (OCB) was poleward of the 42 m radar beam at the beginning of the ESR data. The F region plasma measured by the ESR before 7:05 UT was in the sub-auroral region which was produced by the solar EUV ionization. This region was characterized as smooth (no Ne gradients), cold (low Te), and there was no ion heating events, which indicates no fast plasma flows. After the IMF turned southward, the OCB moved equatorward due to dayside reconnection (Lockwood et al., 1993, 2005; Moen et al., 1995, 2001, 2004a). The OCB crossed the 42 m beam at around 7:05 UT, as is indicated by a sudden enhancement of Te due to cusp electron precipitation (Doe et al., 2001; Moen et al., 2004a). The 42 m beam was located inside the active cusp inflow region during the rest of the time interval shown.

The ionospheric conditions between 7:05 UT and 8:40 UT were characterized by modest electron density (NmF2 of 0.5–2.3 $\times 10^{11}$ m$^{-3}$), enhanced electron temperature which indicates ongoing particle precipitation, enhanced ion temperature due to ion frictional heating which suggests enhanced horizontal ion velocity and/or flow shears, and sporadic ion upflow (positive Vi as shown by red color in Fig. 2g) (Moen et al., 2004a,b, 2006; Skjæveland et al., 2011; Carlson et al., 2012).

As the ESR moved into the postnoon convection cell from around 8:40 UT (see Figs. 1 and 2b), the ESR 42 m observed a series of high density F region plasma enhancements with values up to $5.5 \times 10^{11}$ m$^{-3}$, which were newly produced polar cap patches. These plasma enhancements were originally from

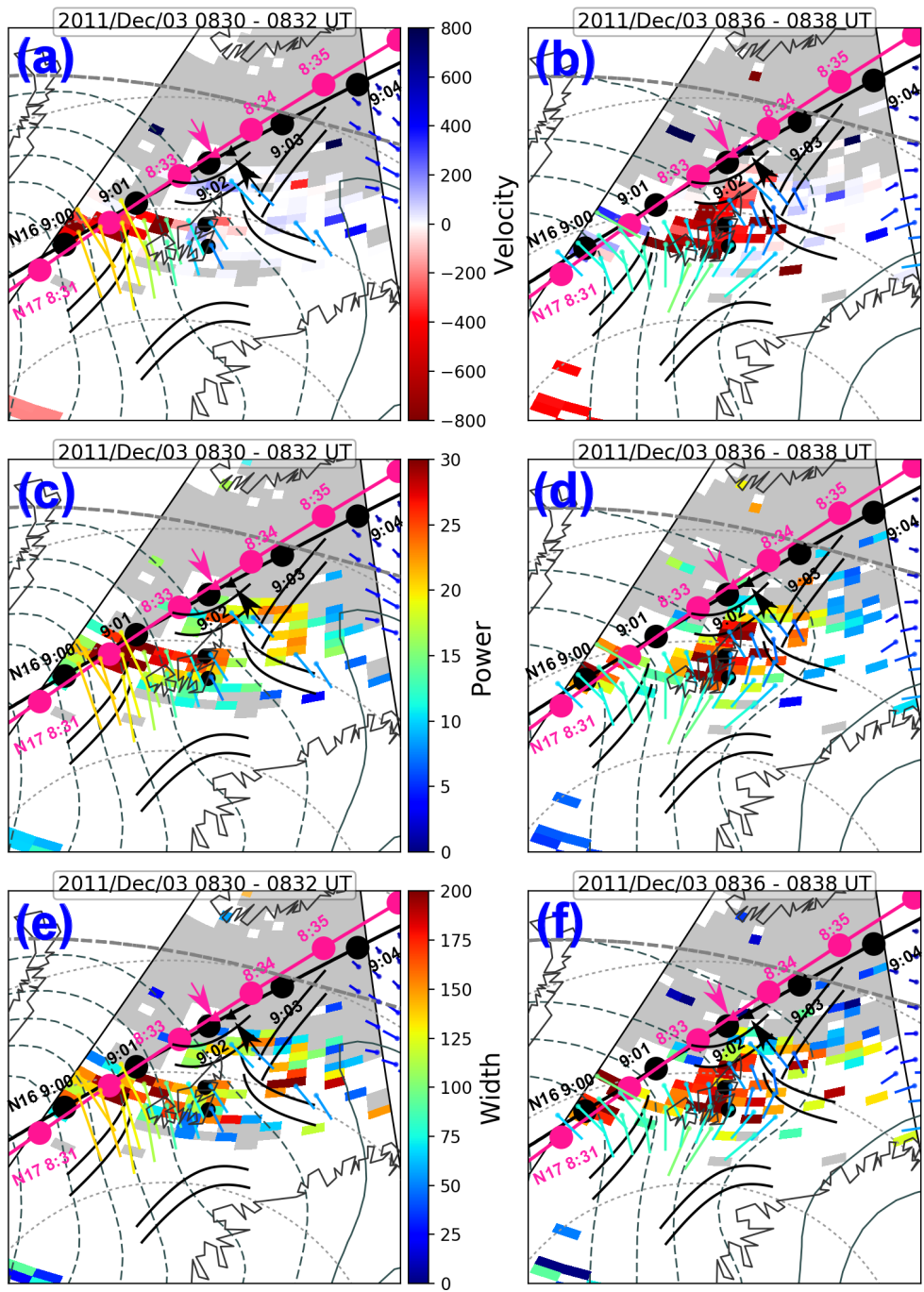

**Fig. 5.** The Doppler velocity (top), backscatter power (middle), and spectral width (bottom) from the Hankasalmi HF radar at 8:30 UT (left column) and 8:36 UT (right column). The ground backscatter, which is characterized by low Doppler velocity and low spectral width, is displayed in grey. The ionospheric convection pattern and fitted drift velocities are also overlayed. The GPS IPPs for the same time periods as Figure 3b are shown as black lines (GPS phase scintillation strengths are not indicated here for clarity). The black and pink lines show the track of NOAA-16 and NOAA-17, respectively. The OCBs identified from both spacecraft are annotated by black and pink arrows and they agree very well, see text and Figure 6 for more detail.

the solar EUV-ionized plasma in the subauroral region from the postnoon sector and were likely sliced by flow shears in the cusp inflow region (see e.g., Carlson, 2012; Moen et al., 2006). The enhanced electron temperature indicates ongoing electron precipitation which was similar to the corresponding prenoon region from 7:05 UT to 8:40 UT. Due to the strong cooling inside the high density polar cap patches (Moen et al., 2004a), enhanced Te was seen near the edges of the polar cap patches. However, the low Te associated with the polar cap patch around 9:10 UT could be a result of less precipitation during this time period. The enhanced Ti signatures were similar to the ones of Te which were also more prevalent at the edges of polar

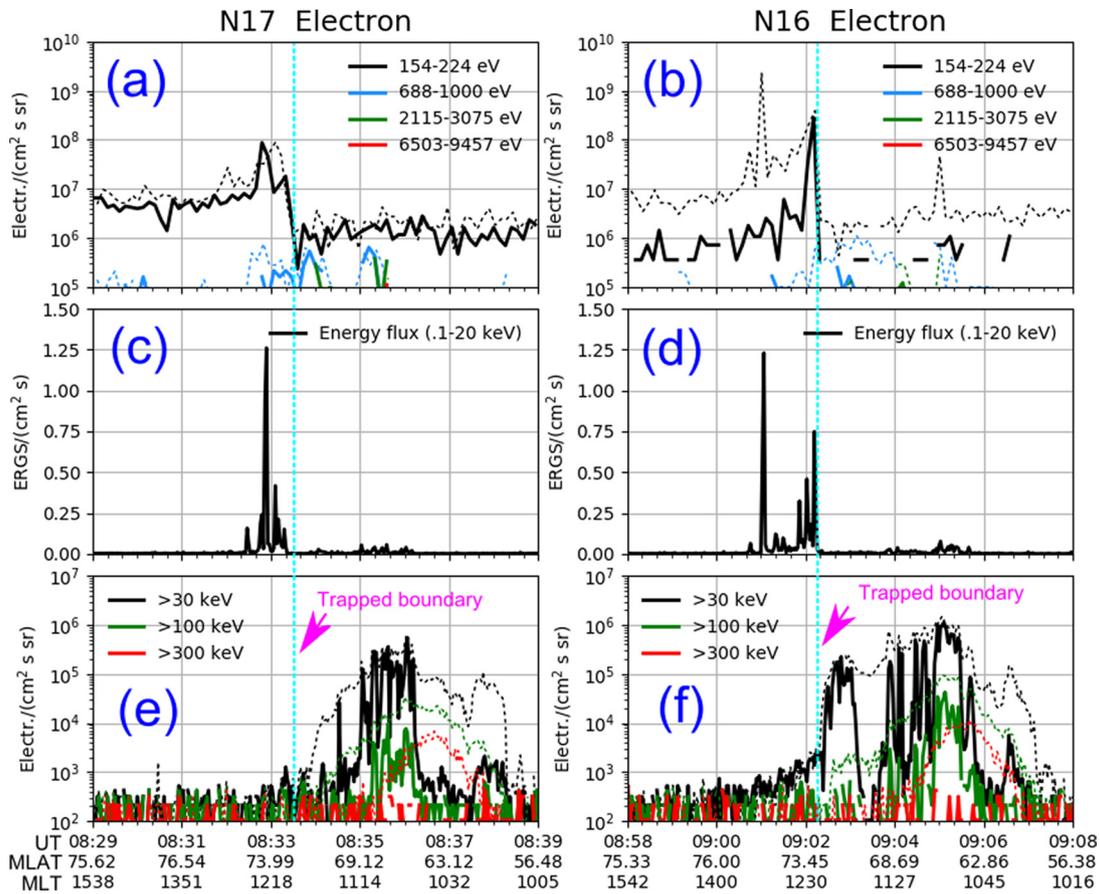

**Fig. 6.** The particle data from NOAA-17 and NOAA-16 spacecraft. In panels a and b, the solid lines show precipitating electron fluxes from four low energy channels in four different colors detected by the 0° TED detector, while the dashed lines show precipitating electron fluxes at 30° from local zenith. Panels c and d show energy fluxes of 0.1–20 keV electrons. Panels e and f show the fluxes of energetic electrons at 3 energy channels from MEPED: the solid lines present the precipitating electrons (10°), and the dashed lines present the trapped electrons (80°). The vertical dashed cyan lines indicate the boundaries of energetic electrons, which is taken as a proxy for the OCB.

cap patches. The ion upflow was similar to that in the prenoon sector.

With the ionospheric conditions presented above, we will now elaborate on the ionospheric scintillation conditions associated with different plasma regimes. Figure 3a and b shows the ionospheric pierce points (IPPs) of selected GPS satellites when Svalbard was in the prenoon and postnoon convection cells, respectively. The GPS phase scintillation levels are presented by different symbols and colors as displayed on the right bottom of Figure 3a and b where red triangles represent $\sigma_\phi > 0.5$ rad. A simple comparison indicates that the scintillation level was higher when Svalbard was in the postnoon convection cell (Fig. 3b), where polar cap patches were formed, than that in the prenoon convection cell (Fig. 3a), where no polar cap patches were formed. Figure 4 is an enlarged version of Figure 3b to allow a more detailed comparison of the relative locations of the ESR and GPS IPPs. The distance between NYA and LYB is about 110 km.

In order to indicate the OCB, we show in Figure 5 the Doppler velocity, backscatter power, and spectral width from the SuperDARN Hankasalmi HF radar at 8:30 UT (left) and 8:36 UT (right). The data were processed using fitex (Ribeiro et al., 2013). The field of view of the Hankasalmi HF radar covers the area of interest. The wide spectral width is a characteristic of the open magnetic flux around magnetic noon (Rodger et al., 1995; Yeoman et al., 1997; Milan et al., 1998; Moen et al., 2001; Chen et al., 2015). From Figure 5e and f, we see that L21 (PRN21 tracked at LYB) was located at the equatorward boundary of a region of broad spectral width, while N21 (PRN21 tracked at NYA) was located inside a region of broad spectral width. The region of broad spectral width is also characterized by high backscatter power and high Doppler velocity.

The black and pink lines in Figure 5 present the trajectories of the NOAA-16 and NOAA-17 spacecraft, respectively. The times are annotated by "HH:MM" in UT near the bold dots. The particle data from TED and MEPED onboard NOAA spacecraft can be used to identify particle boundaries (Moen et al., 1996, 1998; Oksavik et al., 2000). Figure 6 shows the particle data from NOAA-17 and NOAA-16. The poleward boundary of trapped energetic electrons (>30 keV) is well defined in both cases. The boundaries are presented by the vertical cyan lines which delimit a region of trapped energetic electrons (>30 keV) at the equatorward side and soft precipitating electrons (100s eV) at the poleward side. We use these boundaries as a proxy of the OCB and annotate them in Figure 5 by arrows. Figure 5e and f shows that the two boundaries of trapped particles from NOAA-17 and NOAA-16 are consistent, even though NOAA-16 passed ∼30 min later than NOAA-17. The particle boundaries match the boundary of broad spectral width very well. Considering the complexity the HF radio wave propagation (Villain et al., 1985) and the ranging uncertainty (∼115 km by Yeoman et al. (2001)), the boundaries from both techniques are regarded as in a good agreement.

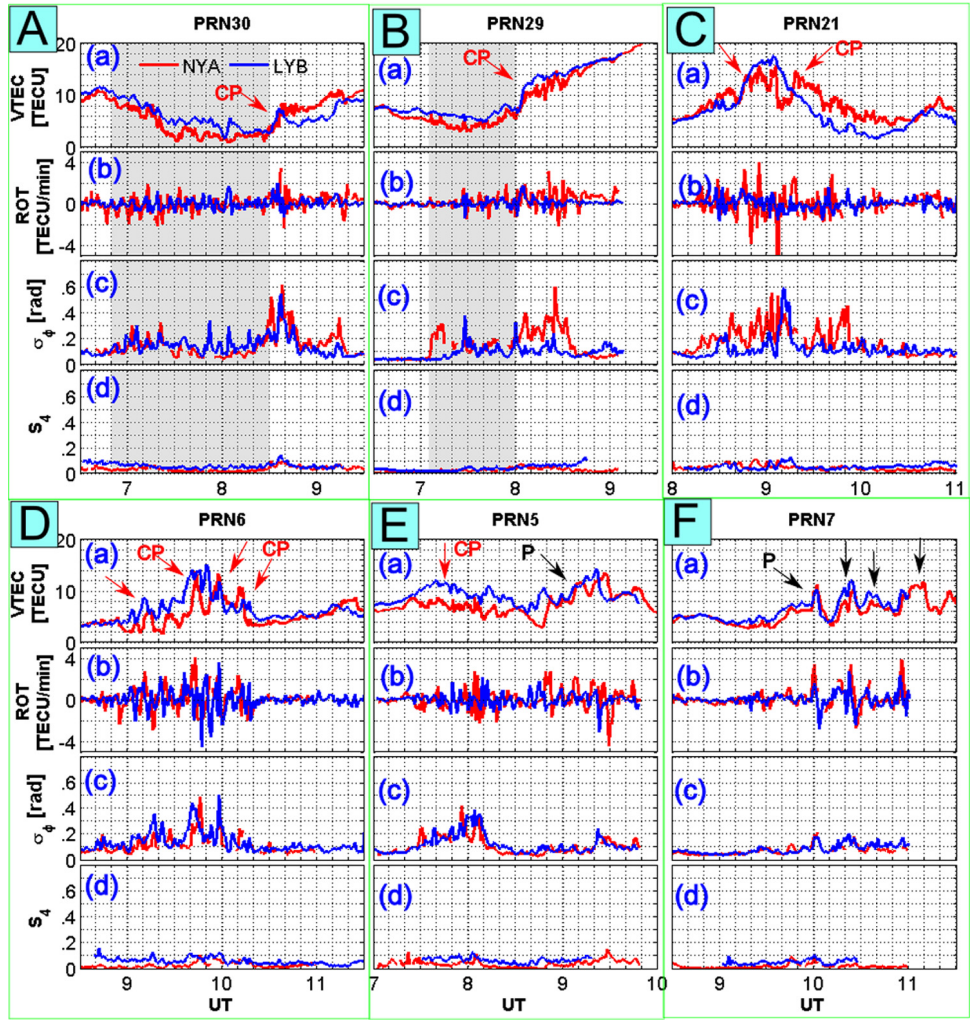

**Fig. 7.** The data from 6 selected GPS satellites as recorded from Ny-Ålesund (in red) and Longyearbyen (in blue). (a-d) The GPS TEC, ROT, phase ($\sigma_\phi$) and amplitude scintillation ($S_4$) indices. The shaded regions in (A) and (B) correspond to the time period when the GPS IPPs were in the cusp auroral region but no polar cap patches were observed. The red arrows indicate the plasma structures of polar cap patches in the cusp auroral region, and the letter "CP" stands for "cusp patch" which means plasma patch in the cusp. The black arrows present the polar cap patches in the dayside polar cap (without particle precipitation), and the letter "P" stands for "patch". See text for more detail.

Figure 7 provides a closer look at the GPS TEC, rate of TEC (ROT), phase ($\sigma_\phi$) and amplitude ($S_4$) scintillation indices for 6 GPS spacecraft. The GPS data recorded at NYA are shown in red, while the GPS data from LYB are shown in blue. During the whole time period, the amplitude scintillations remained very low ($S_4$ below 0.2), while the GPS phase scintillations were very strong and variable. The GPS phase scintillations are produced by electron density irregularities of scale size ranging from hundreds of meters to several kilometers, while the amplitude scintillations are caused by irregularities of scale size from tens of meters to hundreds of meters due to the Fresnel filtering effect (Kintner et al., 2007).

The low amplitude scintillation level is typical for high latitude GPS scintillation measurements.

The shaded regions in Figure 7A and B shows the time interval when PRN29 and PRN30 were inside the prenoon-convection cell with relatively low electron density as compared to later time (Ne from ESR and GPS TEC values). A study of the plasma irregularities observed in this region using in situ measurements from the ICI-3 sounding rocket can be found in Spicher et al. (2016).

The GPS TEC observed by PRN30 decreased from 6:50 UT to 7:30 UT in both NYA and LYB data. This is similar to the electron density observation from the ESR in Figure 2d which also shows a gradual disappearance of the subauroral plasma as the OCB expanded equatorward. The pulsed TEC enhancements were caused by pulsed particle precipitation. The ROT variations between ±2 TECU/min indicate that the electron density inside this region was indeed perturbed as shown from the in-situ rocket observation by Spicher et al. (2016). The GPS phase scintillation was seen up to 0.3 rad, however, the amplitude scintillations were small (below 0.1) even though some minor variations could be seen. PRN29 also observed low and slightly perturbed TEC between 7:05 UT to 8:00 UT. The GPS phase scintillations were up to 0.3 rad before 8:00 UT, however, the amplitude scintillation remained below 0.1. PRN30 (PRN29) observed a polar cap patch from ~8:30 UT (~8:00 UT) as indicated by a sudden increase of the GPS TEC. Then the GPS phase scintillations were enhanced up to 0.6 rad. The amplitude scintillations were also slightly enhanced, but they were still below 0.15.

For PRN21, the data from NYA and LYB were quite different, even though their IPPs were only 110 km away from each other (see Fig. 4). L21 continually observed a smooth increase of GPS TEC from 8:00 UT until 9:05 UT following by a gradually decrease of TEC until 10:10 UT, while N21 showed multiple structures which were characterized by sharp increases and decreases in the GPS TEC data. L21 was located at the equatorward part or equatorward of a region of broad spectral width, where it observed newly arriving plasma from subauroral latitudes (closed magnetic field lines), while N21 observed the newly formed polar cap patches inside the cusp auroral region. These polar cap patches were likely sliced by flow shears when the plasma structures were transported from L21 to N21. N21 observed very dynamic ROT variations between ±4 TECU/min, while the ROT variations of L21 were lower. The GPS phase scintillations of L21 were observed throughout the plasma structure, but $\sigma_\phi$ was higher at the trailing edge (decrease in TEC) of the plasma structure. Similarly the GPS phase scintillations of N21 were also more prominent at the trailing edges of polar cap patches.

PRN6 moved southeastward into the active cusp region (see Fig. 4) and detected a sequence of plasma structures from 9:00 UT to 10:20 UT which are annotated by red arrows and "CP" in Figure 7D. CP is short for "cusp patch" which means the plasma patch in the cusp auroral region. These structures were associated with significant ROT activities and high GPS phase scintillations ($\sigma_\phi$ up to 0.5 rad), while the amplitude scintillation remained low ($S_4 \sim 0.1$).

PRN5 observed large-scale TEC variations and significant phase scintillations ($\sigma_\phi$ up to 0.4 rad) from ~7:20 UT to ~8:20 UT when PRN5 moved from ~74° to ~76° magnetic latitude (MLAT). The variations were probably associated with the formation of a plasma patch in the cusp. As PRN5 continued moving poleward, the TEC variations were even bigger with ROT going to −4 TECU/min. However, the phase scintillations were sporadic and less intense ($\sigma_\phi$ up to 0.25 rad) than before. This TEC variation was probably due to the polar cap patch only (without particle precipitation) according to its location (>76° MLAT). Oksavik et al. (2015) also observed less intense phase scintillations from polar cap patches that were not affected by particle precipitation.

For the scintillation from polar cap patches without cusp aurora, we show data from PRN7 in Figure 7F. PRN7 was located further inside the polar cap (~80° MLAT). The GPS TEC from both stations show variations which were signatures of polar cap patches. The black arrows and letter "P" are used to annotate the polar cap patches in Figure 7Fa. The ROT also shows enhanced variations between ±3 TECU/min. However, the GPS phase scintillations were sporadic and low ($\sigma_\phi$ up to 0.2 rad). This is similar to the observations from PRN5 when it moved beyond 76° MLAT.

## 4 Discussion

We have presented the first direct comparison of the GPS scintillation effects associated with cusp dynamics with and without polar cap patches around magnetic noon. The GPS phase scintillations were at an intermediate level (up to 0.3 rad) in the prenoon sector where there was no formation of polar cap patches. However, when a series of polar cap patches were observed in the postnoon sector, the GPS phase scintillation levels were significantly enhanced (up to 0.6 rad). This result is similar to studies focusing on the nighside (Jin et al., 2014, 2016; van der Meeren et al., 2015), which also show a higher phase scintillation level when the polar cap patches enter into the auroral region and form auroral blobs. This implies that the production mechanisms for plasma irregularities in polar cap patches when they immerse into the auroral region (both dayside and nightside) may be similar as well. In the auroral region, the polar cap patches (or plasma blobs) are subject to auroral dynamics, flow shears and particle impact ionization, etc.

Two main plasma instability modes have been proposed to explain the ionospheric irregularities at high latitudes: (1) the gradient drift instability (GDI) (Keskinen & Ossakow, 1983), and (2) the Kelvin–Helmholtz instability (KHI) (Keskinen et al., 1988). The GDI requires a density gradient and it can produce irregularities at the trailing edge of a plasma patch, while the KHI requires a velocity shear and the irregularities can be created around boundaries of velocity shears, independent of the shear direction. These two instabilities have been under intensive attention in recent literature (see e.g., Oksavik et al., 2011, 2012; Carlson, 2012; Moen et al., 2012; Spicher et al., 2015).

Carlson et al. (2007) proposed a two-step process to explain the rapid onset of plasma irregularities in the cusp inflow region: the KHI first rapidly structures the entering plasma, after which the GDI works on large-scale structures created by the KHI, and produces smaller scale irregularities. From Figure 7Ca, we see obvious differences in the TEC data between two very close IPPs (distance ~110 km, see Fig. 4), which indicates the rapid structuring of the entering plasma from relatively smooth edges (L21) to those with sharp edges (N21). The sharp TEC gradients were likely created by flow shears (see also Fig. 2d for Ne structures from the ESR data). The particle precipitation is unable to produce such large-scale TEC enhancements (>20 min) (Labelle et al., 1989). In addition to the cutting of the entering subauroral plasmas, flow shears can also drive the KHI which creates vortex structures on the flanks of plasma patches (see Fig. 7 of Jin et al. (2016) for more detail). The GDI can work on these sharp density gradients very efficiently (Moen et al., 2012) to produce small-scale irregularities which cause scintillations.

In addition to flow shears in the cusp region (Oksavik et al., 2004, 2005, 2011; Rinne et al., 2007; Moen et al., 2008), the soft electron precipitation could also modulate the entering plasma structure and build up large-scale density gradients (tens of km). Note that Moen et al. (2012) also observed electron density structures of several km scale which were linked to the cusp auroral precipitation. The GDI can work on these gradients to produce smaller scale irregularities (Kelley et al., 1982; Moen et al., 2012; Jin et al., 2015). However, the soft electron precipitation (0.5–1 keV) mainly creates density gradients below 250 km altitude which will disappear in 10 min due to recombination.

Another important question is how active the irregularities are in producing scintillations when the plasma patches move poleward into the dayside polar cap where there are no direct cusp auroral impact. If we assume that the entering high density plasma and the associated irregularities moved poleward into the polar cap along the convection streamlines, we can compare the scintillation levels inside the active cusp region (near PRN21 at ~74° MLAT) and those in the polar cap (e.g., PRN7). PRN7 was located around 80° MLAT, which was 6° poleward of the active cusp. Therefore the TEC variation observed by PRN7 should be due to polar cap patches alone (without cusp aurora), and the GPS phase scintillation was sporadic and low ($\sigma_\phi$ up to 0.2 rad). This is also consistent with observations by Oksavik et al. (2015). By using a drift speed of 500 m/s (Fig. 1b), it takes about 20 min for the entering plasma structures to leave the active cusp region and arrive near the location of PRN7. PRN21 (74° MLAT) detected polar cap patches as early as 8:40 UT, however, PRN7 did not observe clear TEC increase until 9:40 UT. PRN7 might have missed the beginning of the first entering high density plasma and only observed the entering plasma structures at a later time. Another example of the scintillations in relation to the polar cap patch alone can be seen from PRN5. Figure 3a shows that the scintillation level was high when PRN5 was between 74° and 76° MLAT, however, the scintillations became sporadic and low when it moved beyond 76° MLAT (see Figs. 3b and 7E). The observations of PRN5 and PRN7 indicate that the polar cap patches are not associated with significant GPS phase scintillations in the dayside polar cap. This result is consistent with Jin et al. (2015), who shows a significant decrease of the GPS phase scintillation occurrence rate immediately poleward of the cusp auroral region. Clausen et al. (2016) reported a similar result of GPS phase scintillations during periods of sustained dayside and nightside reconnection. They showed that GPS phase scintillations at the dayside are well confined to the cusp region.

When the entering high density plasma structures (patch material) leave the cusp inflow region and move poleward, the ionospheric flow becomes smooth and steady. As a result, the KHI will be less important, and the GDI will be the main operative process. Even though some large-scale density gradients are preserved (see Fig. 7Fa and Fb), the GDI may be less active in producing smaller scale irregularities. Although polar cap patches have been observed to propagate all the way across the polar cap from the dayside to the nightside (Oksavik et al., 2010; Nishimura et al., 2014), poleward of the cusp region, there is a void of intense particle precipitation. As a result, the new density gradients will not be created on the top of polar cap patches. Furthermore, the previous created density gradients by soft particle precipitation in the cusp region disappear as well due to recombination. These impacts altogether make the production of small-scale irregularities to slow down. In the case of a slow production of irregularities, the power of irregularities will be lower than that in the active cusp region. Therefore, the scintillation level is lower in the dayside polar cap than that in the active cusp inflow region. However, enhanced GPS phase scintillation levels are sometimes observed far inside the polar cap (often at the nightside) associated with polar cap patches and large-scale tongue of ionization (Jin et al., 2014; van der Meeren et al., 2014).

## 5 Summary and concluding remarks

In this paper, we have compared the GPS phase scintillation levels in relation to cusp dynamics with and without formations of polar cap patches, and polar cap patches in the dayside polar cap in a case study. The main findings are summarized as follows:

– The highest GPS phase scintillations on the dayside were associated with plasma patches in the active cusp region ($\sigma_\phi$ up to 0.6 rad).
– The plasma structuring can be very rapid in the active cusp region. The GPS IPPs at a distance of only 110 km show obvious different plasma structures.
– The GPS phase scintillations associated with cusp dynamics without polar cap patches were moderate ($\sigma_\phi$ up to 0.3 rad).
– The polar cap patches in the dayside polar cap away from the active cusp were associated with sporadic and moderate GPS phase scintillations ($\sigma_\phi$ up to 0.2 rad).
– The GPS amplitude scintillation level was generally low which is similar to the previous literature.

This study confirms the suggestion of Jin et al. (2015) that the GPS phase scintillation is sensitive to a combination of the cusp auroral dynamics and the intake of the solar EUV-ionized high density plasma. It also agrees well with the study of the nightside auroral blobs (Jin et al., 2014, 2016). It is therefore crucial to take into account the entry of the high density solar EUV-ionized plasma for the future GNSS related space weather products.

*Acknowledgments.* The authors thank the Norwegian Polar Research Institute at Ny-Ålesund for assisting us with the GPS receiver in Ny-Ålesund, Bjørn Lybekk and Espen Trondsen for the instrument operations, Charles Carrano for providing the GPS software, and Vincenzo Romano at INGV for providing the GPS data from Longyearbyen. This research is a part of the 4DSpace Strategic Research Initiative at the University of Oslo. The IMF data are provided by the NASA OMNIWeb service (http://omniweb.gsfc.nasa.gov). EISCAT is an international association supported by research organizations in China (CRIRP), Finland (SA), Japan (NIPR and STEL), Norway (NFR), Sweden (VR), and the United Kingdom (NERC). Data from EISCAT can be obtained from the Madrigal database http://www.eiscat.se/madrigal. SuperDARN is a collection of radars funded by national scientific funding agencies of Australia, Canada, China, France, Japan, South Africa, United Kingdom, and the United States of America. The convection data are retrieved from Virginia Tech servers using the DaViTpy software package. The NOAA particle data were retrieved from the NOAA National Centers for Environmental Information (https://www.ngdc.noaa.gov/stp/satellite/poes/) and were processed by Kjellmar Oksavik (e-mail: kjellmar.oksavik@uib.no). Kjellmar Oksavik is also grateful for being selected as the 2017–2018 Fulbright Arctic Chair, and his sabbatical at Virginia Tech is sponsored by the U.S.-Norway Fulbright Foundation for Educational Exchange. Financial support has been provided to the authors by the Research Council of Norway under Contracts 230935, 230996, 212014, 223252 and 267408. Yaqi Jin wishes to thank the International Space Science Institute in Beijing (ISSI-BJ) for supporting and hosting the meetings of the International Team on "Multiple-instrument observations and simulations of the dynamical processes associated with polar cap patches/aurora and their associated scintillations", during which the discussions leading/contributing to this publication were initiated/held.

The editor thanks Herbert Carlson and an anonymous referee for their assistance in evaluating this paper.